\documentstyle[referee]{mn}
\onecolumn
\input{epsf}
\title{Cluster Winds Blow Along Supercluster Axes}

\author[Novikov et al.]{Dmitri I. Novikov $^1$, Adrian L. Melott$^1$, Brian C.
Wilhite$^1$, Michael Kaufman$^1$, \and Jack O. Burns$^2$, Christopher J.
Miller$^3$ and David J.
Batuski$^3$\\
$^1$Department of Physics and Astronomy, University of Kansas,
Lawrence, Kansas 66045, U.S.A.\\
$^2$Office of Research and Department of Physics and Astronomy, University of
Missouri,
Columbia, Missouri 65211, U.S.A.\\
$^3$Department of Physics and Astronomy, University of Maine,
Orono, Maine 04469-5709, U.S.A.\\}

\pagerange{000,000}
\begin{document}

\maketitle

\begin{abstract} 
Within Abell galaxy clusters containing wide-angle tailed radio sources,
there is evidence of a ``prevailing wind'' which directs the WAT jets.
We study the alignment of WAT jets and nearby clusters to test the idea
that this wind may be a fossil of drainage along large-scale supercluster axes.
We also test this idea with a study of the alignment of WAT jets and supercluster
axes.  Statistical tests indicate no alignment of WAT jets towards nearest
neighbour clusters, but do indicate approximately 98\% confidence in alignment
alignment with the long axis of the supercluster in which the cluster lies.
We find a preferred scale for such superclusters of order 25 Mpc $h^{-1}$.
\end{abstract}

\section{Introduction}

Galaxy clusters are often elongated.
Binggeli's (1982) study of Abell cluster data gave the first
indication that they have a strong tendency towards alignment with
(i.e., their semi-major axes point toward)
other clusters at distances of less than around 30 {\rm h$^{-1}$Mpc}.
West's (1989) study of 48 superclusters also gave clear evidence for alignment
of clusters within superclusters on similar scales.  Simulations also
indicate that cluster axes are aligned with neighbouring clusters (Splinter
{\rm et al.} 1997 and references therein).

It has become recognized within the context of structure formation
in hierarchical clustering (``bottom-up'') due to gravitational
instability that the
large-scale weakly non-linear structure closely follows that produced
in what used to be called the ``top-down'' or ``pancake'' theory (Melott
{\rm  et al.} 1983; Pauls
\& Melott 1995 and references therein; Bond, Kofman, \& Pogosyan 1996).  In
this picture,
most galaxy clusters are formed by
the flow of matter along the sheets and filaments that connect neighbouring
clusters (Shandarin \& Klypin 1984, Colberg {\rm et al.} 1998).

For this reason, merging events are often aligned with these structures.
Mergers inject a velocity anisotropy into the cluster that should
persist for several crossing times. It may be a cause for the
tendency of clusters to point to their neighbours as described above.
The anisotropy is a fossil relic of recent merging events, which can be
seen most clearly in the simulation video (particularly the second
sequence) accompanying Kauffmann \& Melott (1992).

Burns (1998) has reviewed the evidence for persistent winds in the
intra-cluster
medium that may exist as a result of these recent mergers (see also Gomez
{\rm et al.} 1997a, Roettinger, Burns, \& Loken 1996).  Gomez {\rm et al.}
(1997b) showed that there was a highly significant correlation between the
orientation of the semi-major axis of the cluster and the direction of these
winds, as indicated
by the bending of jets from wide-angled tailed (WAT) radio sources in
the clusters.
On the other hand, Ulmer, McMillan, \& Kowalski (1989)
found no orientation of Xray
images toward nearest neighbour clusters.

Although there is evidence of alignment between cluster ellipticity and
neighbour clusters and of alignment of cluster ellipticity with the winds
blowing
WAT jets, there has been
no study of the alignment of WAT jets with neighbour clusters.  It is possible
that the ``prevailing wind'' seen in Abell clusters with WATs may be a remnant
 of drainage along large-scale structure.  If this wind is a fossil of
such drainage, one might expect that it would point either to neighbour
clusters or along supercluster axes.

\section{Procedure}

Images from Gomez {\rm et al.} (1997b), Pinkney {\rm et al.} (1993),
Zhao {\rm et al.} (1989), O'Donoqhue, Owen \& Eilek (1990),
and O'Dea and Owen (1985)
are used in
the determination of the orientation of wide-angle tailed radio source
jets.  These images are overlays of 6-{\rm cm} or 20-{\rm cm} VLA data on
x-ray emission contours in the energy band 0.5--2.0 {\rm keV} from ROSAT PSPC
data.  Another image used
was of data from the Westbrook radio telescope at 6-cm. (Vallee,
Wilson, \& VanDerLaan (1979).
To estimate the direction of the ``wind'' within a cluster, lines are drawn
manually upon the WAT jets, and a bisector drawn for the angle created by
these lines.
Since the clusters are all at a large distance from
Earth, a small angle approximation is used.
The orientation of each bisector in Table 1 is reported with
respect to the horizontal.
An orientation of 0$^{\circ}$ or
360$^{\circ}$ corresponds to a cluster whose
WAT ``points'' to the East on a flattened section of celestial sphere.
To reduce the effect of subjectivity upon the estimated orientation
of the WATs, lines are constructed independently by five
different individuals with no knowledge of the cluster's environment.
As can be seen in Table 1, the random uncertainties in the WAT angle
estimation are relatively small and will have little effect on
questions of alignment.

We do not consider sources for which
orientations were difficult to ascertain for any reason (e.g. extreme
ambiguity of orientation; difficulty in determining location of
one of the WAT jets; poor spatial radio spatial resolution of the more
distant WATs; no apparent wind as evidenced by an opening angle near
180$^{\circ}$).  We also exclude WATs which live in
clusters for which no neighbours exist in our
redshift catalog within 30 Mpc $h^{-1}$,
where $h$ = $H_0/100$ $km$ $s^{-1}Mpc^{-1}$.
(This is sometimes due to lack of redshift information.)
This leaves us with twelve of the seventeen different
WATs that were present in our source studies.

Due to the redshift survey incompleteness,
it is not possible to define a complete sample of clusters for the
WATs in this analysis.
However, given these constraints, we believe that the
remaining sample of WAT clusters and their neighbours is not systematically
biased and should be representative of such clusters and their environs.

\begin{table}
\begin{center}
\begin{tabular}{|l|c|c|}\hline
Cluster & WAT Orientation & Uncertainty \\ \hline\hline

A400 & 124.3 & 6.0 \\ \hline
A562 & 320.2 & 0.2 \\ \hline
A690 & 57.6 & 2.4 \\ \hline
A1446 & 215.0 & 7.6 \\ \hline
A1569 & 203.8 & 2.0 \\ \hline
A1656 & 44.6 & 3.5 \\ \hline
A1940 & 328.6 & 1.4 \\ \hline
A2214 & 248.3 & 9.1 \\ \hline
A2304 & 357.9 & 4.2 \\ \hline
A2306 & 292.5 & 1.7 \\ \hline
A2462 & 288.1 & 1.1 \\ \hline
A2634 & 307.6 & 2.8 \\ \hline
\end{tabular}
\end{center}
\caption{Estimated orientation angles for winds in the WAT clusters
studied.  Note that all angles
are taken with respect to the horizontal and are in degrees.  Uncertainties
represent one standard deviation from estimates by five individuals.}
\end{table}

For each of the twelve WATs, we first search the
Abell cluster catalog to find the nearest neighbour cluster.
Although all
WAT angle bisectors are drawn against a flattened celestial sphere, we find
neighbour clusters in three-space.
The cluster neighbours of each WAT source are determined using
Abell clusters of all richness and distance classes north of
-27$^{\circ}$ declination.
In most cases, only clusters with measured
redshifts are
used. However, approximately 15-20\% of Abell clusters with m$_{10}
 \le
17.0$
do not as yet have measured redshifts, so occasionally the Batuski \&
Burns (1985)
m$_{10} - z$ relation is used. The data for the clusters with observed
redshifts come from a variety of
sources including Struble \& Rood (1987) and Postman, Huchra \& Geller
(1992).
However, the majority of the cluster
redshifts with m$_{10} \ge 16.5$ and $R \ge 1$ were supplied
by the MX Survey and its extension (Slinglend {\rm et al.} 1998;
Miller {\rm et al.}--in preparation).
The MX Survey was designed to
measure all $R \ge 1$ Abell clusters with m$_{10} \le 17.0$ in
the Northern Hemisphere. Currently, the sample of
$R \ge 1$, $0^h \le \alpha \le 24^h$, $-17^{\circ} \le \delta \le
90^{\circ}$,
and $|b| \ge 30^{\circ}$ Abell clusters is
87\% complete to m$_{10} = 17.0$ with 282 out of 324 having measured
redshifts.
Once the $R=0$ clusters are included, the sample is 80\% complete with
457 out of 569 clusters having measured redshifts.
About 90\% of the clusters we use have measured redshifts. Two of the
nearest neighbours (for A562 and for A2306) have estimated redshifts, but
dropping these would not modify our conclusions about nearest neighbours
described in the next section.  The remaining estimated redshifts are
merely a source of foreground/background noise, since we study projected
alignments.

All calculations in this paper are made
using lines projected upon the
celestial sphere and then flattened due to the assumption of a small angle
approximation.  These calculations should be valid, however, since all these
cluster neighbourhoods pictured are relatively small.  The largest angular
separation between WATs and included clusters is approximately
25$^{\circ}$.

Once the nearest neighbour is determined, a
line is constructed which connects the WAT cluster to the nearest neighbour.
The angle $\phi_{\rm i}$ between this line and the WAT bisector is
recorded for each of the
clusters in Table 2 (index {\rm i} denotes the number of the WAT).
Figure 1 shows the cluster environments and the various orientations we
compare.
The distribution of these angles is under
consideration.
We also have available an image which can be found at 
http://kusmos.phsx.ukans.edu/images/A2634SUW.jpg
which shows the cluster gas in an X-ray false color image, the jets from VLA
data, and the cluster oriented in its supercluster environment.

\begin{table}
\begin{center}
\begin{tabular}{|l|c|c|}\hline
WAT Cluster & Nearest Neighbour & Angle \\ \hline\hline

A400 & A397 & 36.0 \\ \hline
A562 & A556 & 3.3 \\ \hline
A690 & A699 & 17.6 \\ \hline
A1446 & A1402 & 22.5 \\ \hline
A1569 & A1526 & 70.3 \\ \hline
A1656 & A1367 & 62.9 \\ \hline
A1940 & A1936 & 37.0 \\ \hline
A2214 & A2213 & 14.0 \\ \hline
A2304 & A2304 & 65.5 \\ \hline
A2306 & A2305 & 69.9 \\ \hline
A2462 & A2459 & 82.2 \\ \hline
A2634 & A2666 & 50.3 \\ \hline
\end{tabular}
\end{center}
\caption{The nearest neighbour cluster and angle between WAT orientation and
a line connecting the WAT and nearest neighbour for each of the clusters
studied.}
\end{table}

\begin{figure}
%\begin{center}
\epsfysize=2.9in
\epsffile{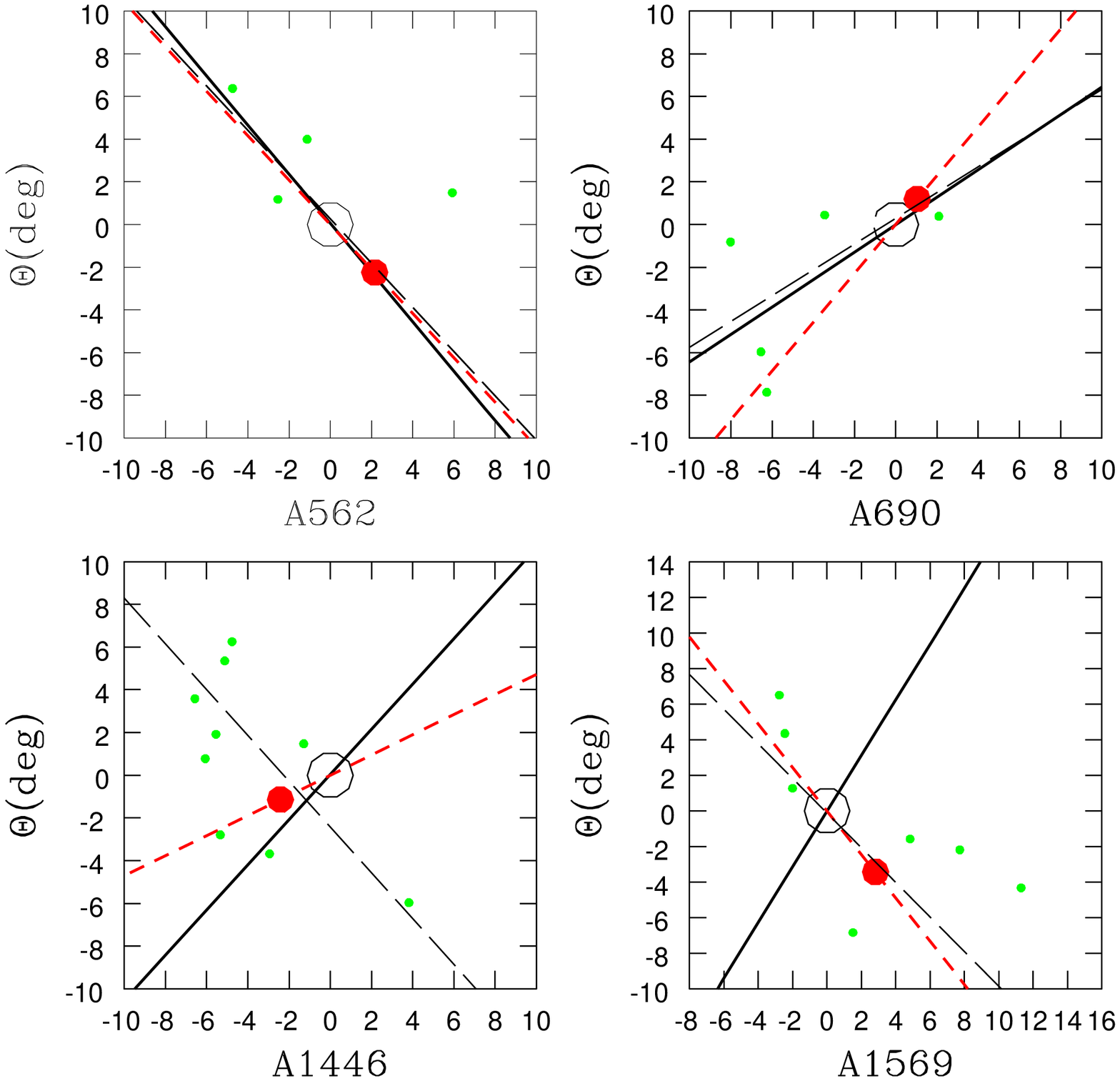}
\epsfysize=2.9in
\epsffile{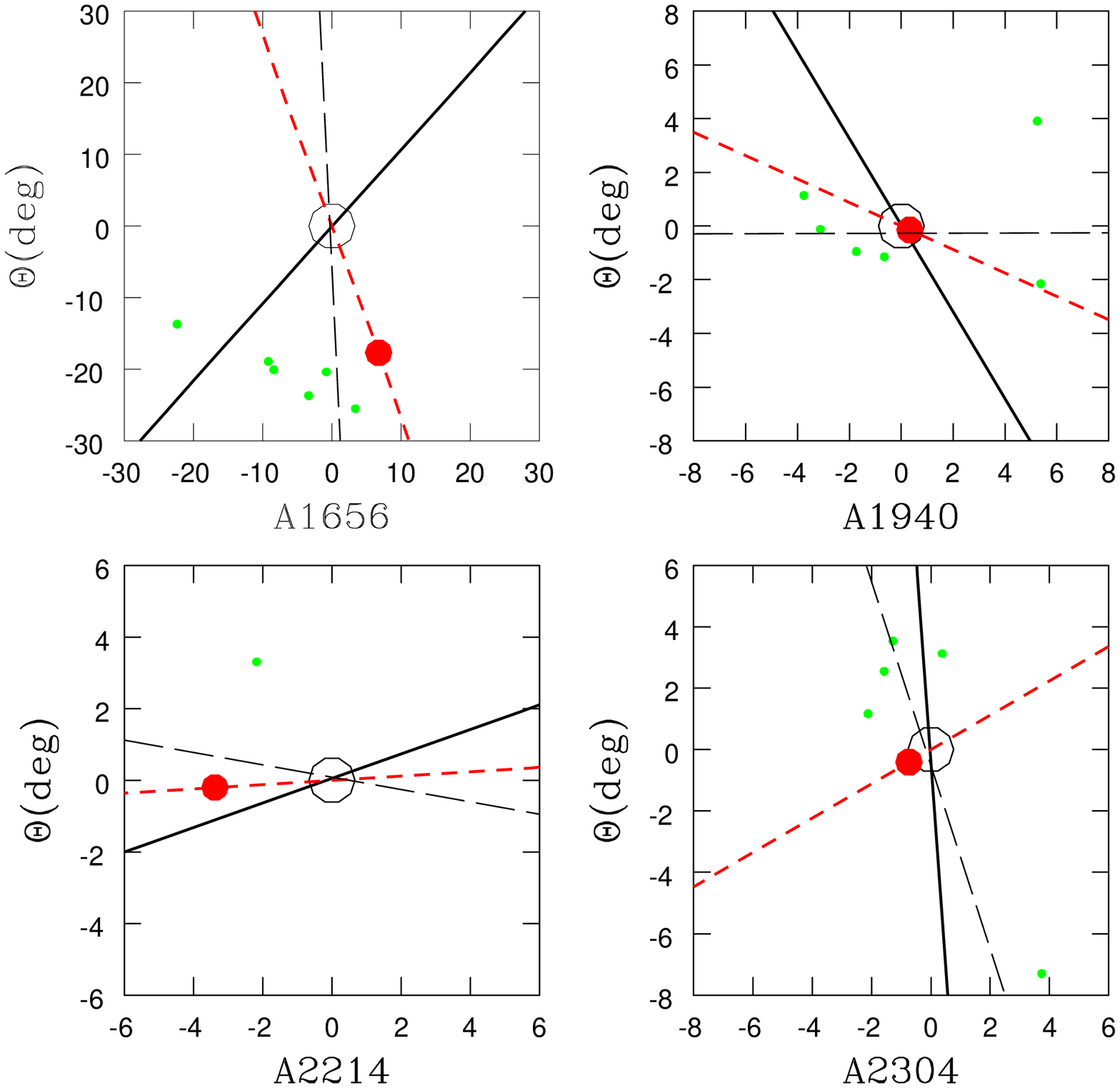}
\epsfysize=2.9in
\epsffile{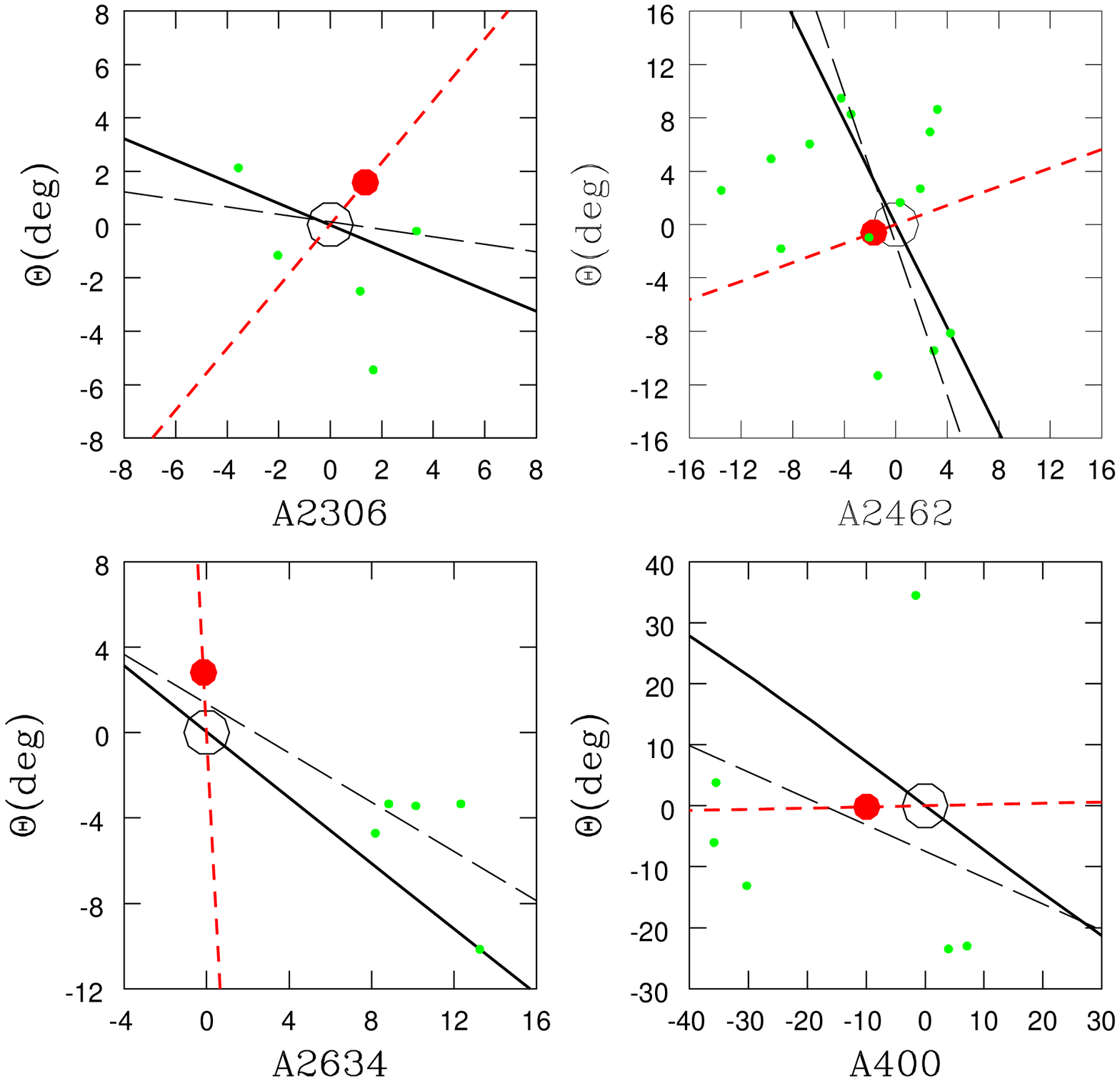}
%\end{center}
\caption{
Open circle: WAT cluster;
solid line: orientation of the wind
blowing the WAT jets;
short-dashed line connects WAT cluster with
nearest neighbour cluster (large darkened circle);
remaining points: clusters
within 50 $Mpc$ $h^{-1}$ of the WAT cluster;
long-dashed line: orientation of supercluster
as described in text. It must be
emphasized that these figures are in projection, while the fit
is weighted by full three-dimensional redshift space distances between
clusters.  Thus the line may not {\it appear} to be a good fit to
the distribution of points.}
\end{figure}

If there is no correlation between the orientations of WATs and
directions of the lines connecting WATs and their neighbours, the angle
$\phi$ between them should be uniformly distributed from 0 to 90$^{\circ}$
and have a mean angle
\begin{equation}
\overline{\phi}=\frac{1}{N}\sum\limits_{i=1}^{N}\phi_{\rm i}=45^{\circ},
\end{equation}
where N is the total number of WATs. If there were any
alignment between WAT bisectors and nearest neighbour clusters, then
$\overline{\phi}$ would obviously be less than 45$^{\circ}$. Equation (1) is
accurate
for large N, but
we have only twelve WATs. $\overline{\phi}$ is 44.4$^{\circ}$,
and a Kolmogorov-Smirnov test shows no significant evidence
that the distribution
is non-uniform (see the next section).

We also check for WAT alignment with the supercluster
in which the cluster lies.
We again search the Abell catalog, this time to
locate clusters within 50 $Mpc$ $h^{-1}$ of each
WAT cluster.  We loosely call such a set of clusters (including the WAT and
nearest neighbour clusters) a ``supercluster.''
We do not include clusters at larger distances from the WAT because this
typically forces part of the volume into galactic obscuration or out of
the survey region.
We have simulated the effect of additional neighbors distributed randomly
with the global sample mean density
out to 100 $Mpc$ $h^{-1}$ and find that, while it adds some extra
noise, it does not remove our alignment signal.

We next determine the orientation of the supercluster long axis.  We  wish to
fit a straight line to the collection of clusters by drawing a least-squares
line based on the projection of each cluster on the flattened celestial
sphere.
We need to include clusters within a finite distance, to take account of
the fact that nearby clusters are more likely to lie within the same
structure as the WAT.  However, we would like to avoid sudden
changes in orientation as this limit is changed to include
a new cluster.  We therefore
apply a least-squares fit to a straight line, but clusters are
Gaussian weighted $\exp (-r^2/2r_0^2)$ for proximity to the WAT cluster.
An advantage of this approach
is that we can explore the effect of changes in $r_0$.
The angles
between the WAT bisectors and these lines are calculated and
recorded in Table 3.
We have used bootstrap resampling (Barrow, Bhavsar, \& Sonoda 1984) to
estimate the uncertainty in the angles.  For our small number of clusters
this procedure is likely to overestimate the uncertainties, so these
can be regarded as upper limits to one-$\sigma$ uncertainties.
It is also true that our result does not depend on the assumption that the
superclusters are straight lines, while these are the uncertainties in a fit to
a straight line. But it is the best we can do to associate some kind of error
bars with the orientation of the supercluster.
The significance of a correlation
between orientation of WAT sources and supercluster axes can
be estimated by investigation of the distribution of these angles in the
same way as for nearest neighbours.

Figure 1 shows the clusters in the vicinity of the named (WAT-bearing)
cluster, along with the orientations of the putative wind (WAT bisector),
the direction to the nearest neighbour cluster, and the axis fit
as described above.  It is important that although actual distances in
redshift space are used to decide the weighting, these Figures are seen in
projection.  Since no radial component of the WAT plasma motion is known,
we can only look for correlation in the projected angles.
Our weighting is based on three-dimensional distances, but the
pictures are in projection,
so the orientation line may not {\it appear} to be a good fit
to the positions of the clusters.

\begin{table}
\begin{center}
\begin{tabular}{|l|c|c|}\hline
WAT Cluster & WAT-Supercluster Angle & Uncertainty \\ \hline\hline

A400 & 11.0 & 24.1 \\ \hline
A562 & 3.5 & 13.2 \\ \hline
A690 & 0.4 & 8.7 \\ \hline
A1446 & 86.2 & 7.6 \\ \hline
A1569 & 77.1 & 10.8 \\ \hline
A1656 & 45.6 & 7.6 \\ \hline
A1940 & 59.6 & 8.5 \\ \hline
A2214 & 27.0 & 23.0 \\ \hline
A2304 & 14.8 & 14.6 \\ \hline
A2306 & 12.4 & 25.1 \\ \hline
A2462 & 6.3 & 12.2 \\ \hline
A2634 & 6.8 & 7.3 \\ \hline
\end{tabular}
\end{center}
\caption{The angle between the WAT jet bisector and the line defining the
orientation of the supercluster for each of the WAT clusters studied.
The supercluster orientation is that for $r_0 = 21 Mpc$ $h^{-1}$ (see text).}
\end{table}

\section{Results and Discussion}

We perform two Kolmogorov-Smirnov (K-S) tests.  The first is done
on our distribution of angles between the WAT angle bisector and the line
connecting the WAT cluster to its nearest neighbour.  This test indicates only
a 1.6\% confidence level that we may reject the hypothesis
of uniform angle distribution.
One may ask if cluster wind directions and nearest neighbour
directions are correlated with cluster axes, why they
are not correlated with one another.  It is not required,
but we would like a physical explanation. We speculate that cluster
axes are affected by both tidal forces and by merger events.
One would expect the nearest cluster to dominate the tidal field,
but merger events should be correlated with the supercluster axis.
Further study is needed to understand this null result.

\begin{figure}
\epsfysize=3.3in
\epsffile{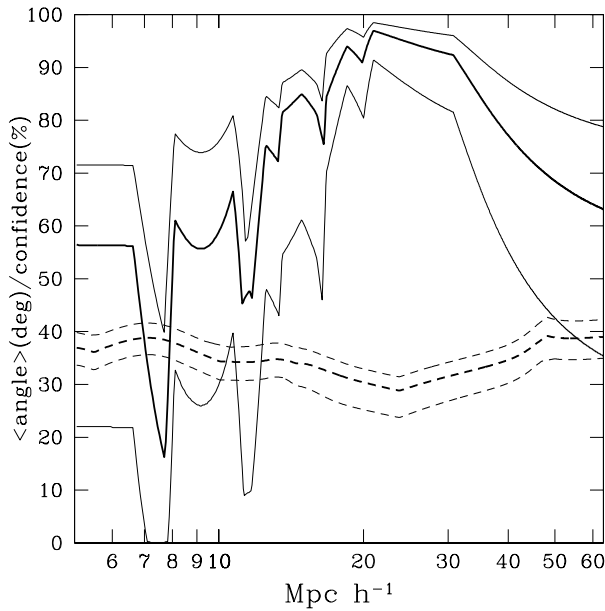}
\caption{A summary of the results of the supercluster-WAT orientation
study.  Results are plotted as a function of $r_0$, the Gaussian smoothing
length for weighting to determine the supercluster orientation.  Larger
$r_0$ corresponds to considering a larger neighbourhood of the WAT.
The heavy dashed line is the mean angle $\overline{\phi}$ between the wind
and the supercluster axis.
The lighter dashed lines are the same mean computed using the eleven best
and eleven worst alignments.  The heavy solid line is the confidence level
(as computed by a K-S test) that the  distribution of angles in the parent
population is not uniform.  The light solid lines are the same confidence
drawn from the eleven best and worst as described above.  It is clear that
for superclusters defined in this way, there is a strong tendency for the
winds to be aligned with the surrounding region on a scale of about
25 $Mpc$ $h^{-1}$.}
\end{figure}

The second test is on the distribution of angles between the
wind (WAT angle bisector) and the supercluster line as defined previously.
The angle and therefore the significance of its
distribution are obviously functions of $r_0$.
In Figure 2, the heavy solid line shows
the confidence level (for rejection of the null hypothesis that the angles
may be distributed uniformly) as a function of $r_0$.  This
confidence reaches a maximum of 97.0\% for $r_0 = 21 Mpc$ $h^{-1}$.
The lighter solid lines are a measure of the uncertainty in
this confidence, generated by choosing eleven best or eleven worst
aligned out of the twelve regions.  There is clearly a preferred
scale, about 25 $Mpc$ $h^{-1}$.  The heavy dashed line is
$\overline{\phi}$, which reaches a value of about 29.3$^{\circ}$,
when the K-S test reaches maximum confidence.
(This is not quite the minimum, which is 28.8$^{\circ}$.)
The mean never exceeds 45$^{\circ}$, although the
confidence is poor for small and large $r_0$.
Discontinuities in the slope of the confidence curve occur when the slowly
rotating supercluster orientations cross 0$^{\circ}$or 90$^{\circ}$.

For small $r_0$, the supercluster orientation
is poorly determined due to the small number of objects included.
As $r_0$ approaches zero, this becomes the nearest neighbour
test.  For large
$r_0$, we exceed the scale of typical segments of the network.
It is interesting that the region of best $r_0$ coincides with the wavelength
of perturbations going non-linear today as estimated elsewhere (Melott and
Shandarin 1993).
It is also close to the neighbour linkage radius found
necessary for percolation in one supercluster study (Batuski
{\rm et al.} (1998).  We see here support for the idea that the structures
found by percolation have a dynamical origin, and are not merely accidental
artifacts.

\begin{figure}
\epsfysize=3.3in
\epsffile{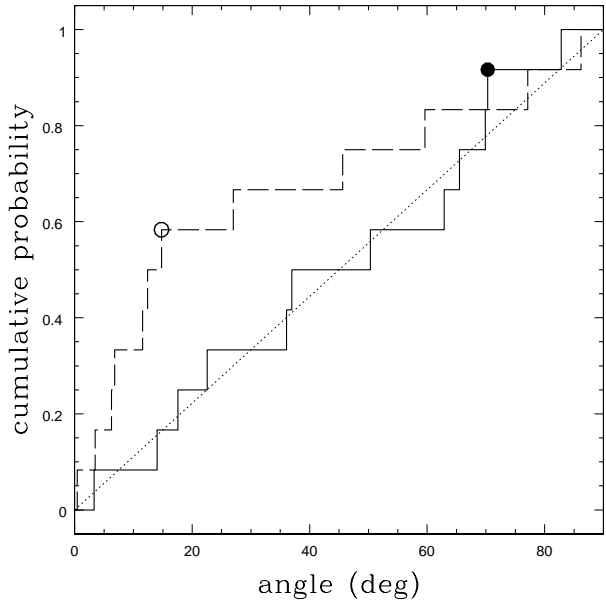}
\caption{A graphical representation of the Kolmogorov-Smirnov tests performed.
The dotted diagonal line represents the expected relationship between angle
measures and cumulative probability.  The solid stair-step pattern represents
this relationship within the nearest-neighbour distribution of angles.  The
dashed stair-step pattern corresponds to this same relationship for the
distribution of supercluster angles.  The larger the maximum deviation of the
stair-step pattern from the expected diagonal line, the higher the confidence
with which one can reject the hypothesis of a uniform distribution of angles.}
\end{figure}

The first K-S test result is consistent with the WAT
orientations being
completely uncorrelated with nearest neighbours.  However, the second K-S test
result indicates a strong correlation between WAT wind orientation and the
supercluster orientation.
It should be noted that a sample size of twelve WAT clusters is very small.
West (1989), for example, studied 48 clusters.
We recommend investigation of a larger data sample.
This will be a lengthy process, dealing with a large number of unclassified
sources in the literature.

However, we believe our conclusions on the supercluster-WAT alignment are
robust,
and offer several reasons
for this.  First, the K-S test is useful under certain conditions for small
sample sizes.  It is statistically robust at the sample size and confidence
level of our result (Lehmann and D'Abrera 1975).

Another approach is to use a completely different statistic. If we put
the angles into four bins of 22.5$^{\circ}$, each of which would be
equiprobable
under a uniform population, we can use the binomial theorem to evaluate
the probability that seven or more out of twelve will lie in the first
bin.  The result is 98.9\% confidence that the distribution is not
uniform.  Using three bins of 30$^{\circ}$ produces a similar result (98.5\%).
Apparently the effect we have found is so strong it is significant even for
a small sample size.

Our results show a correlation between objects two orders of magnitude
apart in size. They are also dynamical evidence in favor of quasi-linear
hierarchical clustering following ``pancake'' dynamics
(Melott and Shandarin 1993).  This has had great success in
reproducing large structure in N-body simulations.
It predicted the
supercluster-void picture of large-scale structure accepted today
(Zel'dovich, Einasto, \& Shandarin 1982, Melott {\rm et al.} 1983).
However, this general agreement between theory and observation
has been based on statistical measures of the galaxy
distribution, not on observed dynamics.  Analysis of cosmic flows has not yet
progressed to showing features unique to the quasi-linear regime.  The
flows indicated here are not a part of linear theory, and thus lend
empirical support to the quasi-linear analysis of gravitational
instability.

\section{acknowledgments}
We are grateful to the referee, M. West, and also to S. Bhavsar for helpful
comments and conversations.
DN was supported by an NSF-NATO Fellowship and other authors at the University
of Kansas were supported by the NSF-EPSCoR program.
JOB  was supported by the NSF.
CM was funded in part by NASA-EPSCoR program
through the Maine Science and Technology Foundation.

\end{document}